# Literature Review:

## Molecular Communication Systems Design for Future City

**First Year PhD Progress Document**

**School of Engineering**

**1256827**

**01/02/2014**

# *LITERATURE REVIEW*

## 1. INTRODUCTION

An area of interest in the modern age is the human migration from rural areas to cities. Cities are characterized by a dense concentration of buildings and key infrastructures. However, what has been lacking is a pervasive sensor technology that can monitor the performance of these structures. Partly, this is due to the fact that the information collected from sensors cannot be easily transported from the embedded location to an external data hub. Examples of health monitoring in structures include monitoring corrosion, fracture stress, and material delamination. The scenario examples include: pipelines, tunnel networks, and some industrial and medical areas such as hospitals (minimise electromagnetic interference), and turbines.

Whilst electromagnetic (EM) waves are widely used for wireless communications, there are significant challenges in applying outdoor communication systems to indoor and embedded areas. The key challenges include the ratio of the antenna size [1], the propagation environment, in particular the penetration loss suffered from building materials. However, in recent years, scientists and engineers have proposed alternative ways to achieve wireless communications.

Inspired by nature, there is an entirely different method of communication which could be considered to solve the problems. Molecular Communication [2], [3] uses special molecules which generally are chemical signals to carry information for communication. The basic concept of the molecular communication system is similar to the radio wireless communication systems. First the information is modulated into different properties such as number [4], concentration [5], ratio of molecules [6] and release time [7]. Then the emitter could release the information-carrying molecules into the transmission channel where these molecules propagate through variety ways such as molecular motors [8], air flow [9], molecular diffusion [10] and bacteria [11] until they reach the receiver. At last the receiver could detect the molecules and extract the information from these molecules.

As mentioned at the beginning, this literature review will focus on the macro-scale molecular communication. The advantages of molecular communication will be analysed, the theoretical model of macro communication system will be presented and finally explain the feasibility of the practical system construction.

## 2. MOLECULAR COMMUNICATION ADVANTAGES

Molecular communication worth to be researched is due to the inherent advantages over traditional communications scenarios.

The most notable advantage is scalability and energy efficiency [12]. For example, the molecular communication is naturally used at nano- and micro- scale for intra/inter-cellular communications [13], and be applied as pheromones at macro- scale between the same species [14].  Compare to the EM wave based communication system, molecular systems consume much less energy [12]. Moreover, the primary advantage of molecular communications is that it is not subject to the EM wave laws of diffraction and propagation. Micro- and nano-sized molecules can navigate a variety of environments with ease. Furthermore, the transmitter size is not restricted to be proportional to the EM wavelength, and the potential for ultra-low energy emission is existent. Finally, Molecular communication can be bio- and chemical- compatible [15].

## 3. THEORETICAL COMMUNICATION MEDEL

Despite all the recent work in chemical signalling, there have been few practical demonstrations of molecular communication systems that can be used to transfer generic messages in a continuous and reliable manner. One of the major obstacles in implementing molecular communication is the tedious, laborious and expensive nature of wet lab experimentation. Another challenge is the difficult nature of modulating generic signals onto chemical compounds, and emitting it in such a way that a reliable reception can be achieved.

As a result, a large body of work on the theoretical aspects of microscopic molecular communication systems has been developed [16]–[19], without any physical implementation of a fully functional communication device. The theoretical framework reveals interesting communication bounds for conveying data using nano-particles. In this subsection, the basis for a theoretical understanding of the communication limits, and how it is applicable to real system implementation will be discussed.

1) *Random Walk - Pulse Response:* Let us assume that the information bearing nano-particles undergoes a random diffusion process (random walk). Consider an emitter that emits a single pulse of chemicals because generic communications system will modulate messages into a succession of individual pulses.

At time $t$ after an emission, the probability density function of the molecule concentration at any point $x$ away from the point of emission follows an inverse exponential function [20]:

$$f(x,t) = \frac{1}{\sqrt{4D\pi t}} exp\left(-\frac{x^2}{4Dt}\right). \quad (1)$$

for a given diffusivity $D$, which is a chemical-medium dependent measure of the rate of diffusion.

In order to capture the molecules at the receiver, the probability of capture is [19]:

$$p_c(x,t) = erfc\left(-\frac{x}{2\sqrt{Dt}}\right). \quad (2)$$

For intra-cellular chemical signalling, the diffusivity $D$ value is $1 - 300\mu m/s^2$, and the diffusion distance $x$ is $1–200\mu m$. In such a scenario, the probability of capturing 90% of more of the emitted molecules can be achieved in less than a milli-second.

For inter-organism chemical signalling, the diffusivity $D$ value is $0.1 - 1cm/s^2$, and the diffusion distance $x$ is several metres. In such a scenario, the probability of capturing 90% of more of the emitted molecules can be achieved in a few minutes to an hour.

In reality, the diffusion process is assisted by currents both inside the body and between bodies. Air currents have an effect of rapidly accelerating the diffusion process and hence, chemical communications over several metres is capable at speeds in the order of seconds to a minute.

2) *Pulse Modulated Signal:* Given the pulse response in (2), a sequence of pulses can be examined. A key criterion for reliable detection of continuous pulses is to avoid overlapping pulses at the receiver, such that the response of one pulse overly interferes with the shape of another. This is known as inter-symbol-interference (ISI).

In EM-based signalling, digital filters are used to shape the transmit pulses so that ISI is minimized. Fortunately for EM waves this is possible, as the time gap between pulses (milli-seconds) is much greater than the stochastic nature of the channel (nano-seconds). However, in chemical signalling the channel is extremely stochastic, and the delay spread of the channel can be in the order of seconds to minutes. This potentially means, in order to avoid excessive ISI, the time separation between successive pulses need to be in the order of seconds to minutes as well. As mentioned, the delay spread from multiple EM-waves reflecting off surfaces is small (10–50 nano-seconds in urban environments over several kilometres) compared to the guard time ($T$). This is primarily due to the speed of light and the absorbing nature of materials such that the energy from multiple reflections is often lost. However, for molecular diffusion, this is not the case. The delay spread can be several seconds over just a few metres of space. An analogy would be the scent or smell that lingers indefinitely.

*3) Data Rate Scalability:* The resulting consequence of strong ISI in the chemical diffusion channel is that in order to achieve a reasonably reliable communication link, the channel capacity is low (in the order of 0.1 bits/s per molecule type [19]). This may sound tremendously underwhelming, when considering that a modern Wi-Fi or 4G cellular link can provide up to 100 Mbits/s. Modern EM-based systems use the bandwidth resource to scale up its data rate. The data rate $R$ is given by the product of the bandwidth $B$ and the capacity $C$, whereby the capacity is a function of the channel quality $S$:

$$R\left[\frac{bits}{s}\right] = B[resource] \times C(S)\left[\frac{\frac{bits}{s}}{resource}\right]. \ (3)$$

The typically quoted $R$ =100 Mbits/s in modern communication systems is spread over a $B$ =20 MHz channel, yielding a real channel capacity of $C$ =5 bits/s per unit frequency (Hz). This is achieved only when the signal power of the channel is $S$ =1000 (30dB) times higher than the noise power. Furthermore, this is only realizable with state of the art channel modulation and coding schemes, which have been developed over a period of 50 years.

## 4. PRACTICAL SYSTEM CONSTRUCTION

The idea of the nano-particle communications test-bed was conceived in summer 2012 by Dr. Weisi Guo. In the article [21], the system fills an important gap in the molecular communication literature, where much current work is done in simulation with simplified system models. The demonstrator test-bed is a first-generation device that the team hopes will kick-start academic and industrial revolutions in designing molecular-based communication systems. The system is inexpensive to build, and the platform is available for sale as a modifiable and re-programmable research test-bed. The test-bed primarily consists of a transmitter and a receiver. The propagation channel in between is several metres of either free-space or a specific structural environment (i.e., a tunnel network). On the transmission side, the hardware consists of: i) a user inter-face for text entry, ii) a micro-controller that converts the input text into binary sequences and then modulates the sequence on chemical signals, iii) a reservoir of chemicals, and iv) a chemical release mechanism. On the receiver side, the hardware consists of: i) a chemical sensor, ii) a micro-controller that demodulates and decodes. The type of data we demonstrated is a short string of text data, because text-based information is of interest to sensor networks and command-based communication systems. For example, the team members are exploring the possibility of using molecular communications for structural health monitoring and giving commands to underground robots. A key finding is the nonlinearity of the platform. This finding is very important because most of current communication theory is based on linear systems. If it is shown that the nonlinearity is part of practical molecular communication systems (i.e. the nonlinearity cannot be resolved using better equipment), new communication theoretic work may be necessary on this topic.

The comparison of the modern EM-based 4G Long Term Evolution system and Kinboshi molecular communication system is shown in Table 1.

TABLE 1 COMPARING ELECTROMAGNETIC WITH CHEMICAL COMMUNICATIONS

| Parameter | EM | Chemical |
|---|---|---|
| System | 4G LTE | Kinboshi |
| Resource | Bandwidth (20MHz) | Chemical Types |
| Range | Very Long (km) | Short (m) |
| Delay Spread | Small (ns) | Long (s) |
| Reliability | Very High | Medium |
| Peak Capacity | 5 bits/s/Hz | 0.3 bits/s/chemical |
| Emitter Size Limitation | $\propto$ Wavelength (mm-cm) | >Molecule Size |
| Propagation Law | Maxwell | Brownian Motion |
| Artificial Gain | Antenna Gain | Drift Current |
| Emission Type | Active (Antenna) | Passive or Active |
| Energy Consumption | High ($\approx Watts$) | None (Passive) or Low (Active) |

## 5. CONCLUSION

This literature review offers a brief explanation of the molecular communication. Firstly the review introduced the motivation on researching on molecular communication and went on analysis the advantages and challenges molecular communication has. Then the first practical molecular text message system was introduced followed by proposing theoretical models of molecular communication systems. Finally the comparison of the existing EM communication system and Molecular communication system was summarized. The need to convey information has always existed in both the animal and human kingdoms. There is now an increased focus to communicate in extreme environments, such as between micro-robots performing targeted drug delivery, and between embedded sensors in industrial infrastructures. Therefore, in conclusion, Molecular communication is worth researching in special areas where EM communication is not feasible.